# Analytical Models of the Performance of IEEE 802.11p Vehicle to Vehicle Communications

Miguel Sepulcre, Manuel Gonzalez-Martín, Javier Gozalvez,
Rafael Molina-Masegosa, Baldomero Coll-Perales

*Abstract*— The critical nature of vehicular communications requires their extensive testing and evaluation. Analytical models can represent an attractive and cost-effective approach for such evaluation if they can adequately model all underlying effects that impact the performance of vehicular communications. Several analytical models have been proposed to date to model vehicular communications based on the IEEE 802.11p (or DSRC) standard. However, existing models normally model in detail the MAC (Medium Access Control), and generally simplify the propagation and interference effects. This reduces their value as an alternative to evaluate the performance of vehicular communications. This paper addresses this gap, and presents new analytical models that accurately model the performance of vehicle-to-vehicle communications based on the IEEE 802.11p standard. The models jointly account for a detailed modeling of the propagation and interference effects, as well as the impact of the hidden terminal problem. The model quantifies the PDR (Packet Delivery Ratio) as a function of the distance between transmitter and receiver. The paper also presents new analytical models to quantify the probability of the four different types of packet errors in IEEE 802.11p. In addition, the paper presents the first analytical model capable to accurately estimate the Channel Busy Ratio (CBR) metric even under high channel load levels. All the analytical models are validated by means of simulation for a wide range of parameters, including traffic densities, packet transmission frequencies, transmission power levels, data rates and packet sizes. An implementation of the models is provided openly to facilitate their use by the community.

## I. INTRODUCTION

V2X (Vehicle to Everything) communications improve the safety and traffic efficiency of connected and automated vehicles thanks to the wireless exchange of information. The V2X standards DSRC (Dedicated Short Range Communications) and ITS-G5 (specified in Europe by ETSI) use IEEE 802.11p technology [1] for vehicles to wirelessly exchange messages. IEEE 802.11p is an amendment of the IEEE 802.11 standard designed for vehicular communications.

The critical nature of V2X communications requires their extensive testing and evaluation. This can be done analytically, through simulations and prototyping. Simulations require a high computing power and prototyping consume a high number of human and economic resources. Analytical models provide a reasonable balance between accuracy, computational complexity and scalability. Most of the existing analytical models are based on the well-known Markov model proposed by Bianchi [2] for IEEE 802.11 networks, and usually focus on the detailed modeling of the MAC (Medium Access Control). However, they often simplify the radio propagation conditions and the effect of interferences, or do not model the hidden terminal effect. In fact, existing models frequently assume ideal (a.k.a. perfect) radio propagation conditions or consider a fixed channel error probability. Few studies consider the effects of pathloss and shadowing on the received signal power, or the multipath fading effect on the probability of successfully receiving a packet. The effect of interferences is normally limited to the identification of packet collisions. Thus, all packets received with collision are lost independently of the received signal power of the interfering packet. Some studies model the hidden terminal effect but do not include an adequate and accurate modeling of the propagation and interference conditions. In fact, to the authors' knowledge, none of the existing analytical models accurately model together the propagation, interference and hidden terminal effects. Moreover, existing analytical models provide performance metrics (e.g., PDR or Packet Delivery Ratio, throughput, delay, etc.) generally representing the average network performance. In this case, models cannot adequately account for the impact of the distance between transmitter and receiver and the operating conditions (e.g., the traffic density) on the performance of V2V communications.

This paper progresses the state of the art with a new analytical model of the performance of IEEE 802.11p V2X communications that includes a detailed modeling of the propagation and interference effects, as well as the impact of the hidden terminal problem. The propagation effects include pathloss, shadowing and multipath fading. The received signal power of interfering packets is taken into account to determine if a packet is lost due to a collision or not. Moreover, the

This work was supported in part by the *Generalitat Valenciana* through projects GV/2021/044 and AICO/2018/095, the Spanish Ministry of Science, Innovation and Universities, AEI, and FEDER funds under TEC2014-57146-R and TEC2017-88612-R, and research grant PEJ-2014-A33622. Authors are with Universidad Miguel Hernandez de Elche (UMH), Spain. Contact emails: msepulcre@umh.es, j.gozalvez@umh.es

performance is modeled as a function of the distance between transmitter and receiver, which is important in vehicular networks to identify the distances and conditions under which the requirements of V2X applications can be satisfied.

The proposed model can accurately estimate the V2X communications performance for a wide range of scenarios and operating conditions, and represents an attractive and lightweight alternative to the computationally costly network level simulations. The model can be integrated in road traffic simulators and account for accurate V2X communications without having to implement the complete communications protocol stack or interface with network simulators with the consequent reduction of the simulation time [3][4].

In addition to the PDR model, we present in this paper new analytical models to quantify the probability of the four types of possible packet errors in IEEE 802.11p: errors due to a received signal power below the sensing power threshold, errors due to propagation effects, errors produced because the receiver is busy decoding another packet, and errors due to packet collisions. These probabilities depend on the distance between transmitter and receiver, and are used to estimate the PDR. They also represent a valuable tool to understand the operation of IEEE 802.11p, and design and test mechanisms to overcome some of its limitations. These probabilities are estimated using a new analytical model of the Channel Busy Ratio (CBR) metric that is also presented in this paper. The CBR is utilized to estimate the channel load in vehicular networks. It represents the fraction of time that the channel is sensed as busy, and is an important and widely used metric to quantify the channel load in vehicular networks. Existing CBR analytical models fail to accurately estimate the CBR under high channel loads since they do not consider the compression factor. This factor reduces the CBR experienced when packets collide and overlap in time. To the authors' knowledge, the analytical CBR model presented in this paper is the first to account for such compression, and can hence accurately model the CBR under all channel load scenarios. All analytical models are validated using Veins simulator for a wide set of parameters and an implementation is openly available in [5].

## II. STATE OF THE ART

Different analytical models have been proposed in the literature for vehicular networks based on IEEE 802.11p. A detailed review and analysis of the most relevant models is presented in this paper and summarized in Table I. Table I classifies models based on: 1) communications (unicast -U- or broadcast -B); 2) modelling of hidden terminal effect; 3), accurate modelling of propagation errors; 4) impact of received signal power of interfering packets on packet losses due to collisions. Most of the existing IEEE 802.11p analytical models focus on the modeling of the MAC in order to analyze and optimize its performance. Most of these models simplify the modelling of the propagation and interference conditions, and consider communications without errors within a pre-defined radius $R$ when there is no interference. This simplified modelling of propagation effects result in that packet losses at distances below $R$ can only be produced due to packet collisions, while packets cannot be received beyond $R$. Moreover, the interference power is not taken into account to evaluate if a packet is lost due to a collision or not. In addition, most of the existing studies do not consider the hidden terminal problem, which is particularly relevant in vehicular networks. Simplifying the propagation and interference modeling can significantly impact the outcome and validity of V2X studies given the safety-critical nature of vehicular applications and the scalability challenges of vehicular networks.

A significant number of existing analytical models focus on unicast communications ([6]-[15] in Table I) with typically an Access Point or RSU (Road Side Unit) (e.g. [7] and [9]). Unicast packets require the transmission of ACKs, retransmissions and the adaptation of the contention window of IEEE 802.11p. The vast majority of standardized V2X messages are transmitted in broadcast mode and thus these models are not directly applicable to study the performance of vehicular communications. Although not designed for vehicular communications, the model in [15] is one of the few models that considers propagation errors, but using a simplified model that considers a constant error probability. This probability depends on a pre-configured SNR (Signal to Interference) since all the links are considered to have the same deterministic pathloss. Moreover, it takes into account the received signal power of interfering packets assuming Rayleigh fading. However, it does not take into account hidden terminals. From Table I, none of the existing unicast models jointly considers the hidden terminal problem, the propagation effects and the received signal power of the interfering packets.

The analytical models for broadcast communications ([16]-[28] in Table I) are more appropriate to study the V2X communications performance. Several of these models ([16]-[20]) have been derived using simplified propagation and interference conditions and do not model the hidden terminal problem. For example, the work in [16] proposes a one-dimensional Markov model that modifies the classic two-dimensional Markov model proposed by Bianchi [2] assuming perfect communications, i.e., that there are no packet losses up to a pre-defined distance in absence of interferences. The authors also assume that all vehicles can communicate with each other (i.e., no hidden vehicles) and that a packet collision always results in a packet loss independently of the received signal power. The authors consider a discrete time D/M/1 queue to model the periodic broadcast transmissions from each

TABLE I. COMPARISON OF ANALYTICAL MODELS

| Reference | Mode | Hidden terminal | Propagation errors | Interference |
|---|---|---|---|---|
| [6]-[9] | U | × | × | × |
| [10][11] | U | Yes | × | × |
| [12]-[14] | U | × | FP | × |
| [15] | U | × | Fading | OC |
| [16]-[20] | B | × | × | × |
| [21]-[24] | B | Yes | × | × |
| [25][26] | B | Yes | FP | × |
| [27] | B | × | Yes | Yes |
| [28] | B | Yes | × | OH |
| Our model | B | Yes | Yes | Yes |

Legend. B: broadcast; U: Unicast; FP: Fixed probability; OC: only concurrent transmissions; OH: only hidden nodes; ×: not included.

vehicle. They then use the model to compute the probability of packet collision and the channel access delay as a function of the traffic density. A similar approach is adopted in [17], where authors propose a model based on Markov chains, but without taking into account the hidden terminal effect, and considering simplified propagation and interference conditions. This model was extended in [18] with EDCA (Enhanced Distributed Channel Access) priorities. The model proposed in [19] evolves the existing ones by explicitly accounting for the IEEE 802.11p channel switching. The model proposed in [20] integrates the queuing process and the IEEE 802.11p MAC into a two-dimensional Markov chain. All models in [16]-[20] model in detail the MAC of IEEE802.11p. However, they consider simplified assumptions related to propagation, interference and hidden terminal effects that hinders their capacity to accurately model V2X communications.

The models in [21]-[24] consider the effect of hidden terminals on the performance of V2X communications. In particular, the work in [21] proposes an analytical model to quantify the PDR and the channel access delay in IEEE 802.11p under different traffic densities assuming ideal propagation and interference conditions. To this aim, each vehicle is modeled as an M/G/1 queue with an infinite transmit buffer size, i.e., no packet loss due to buffer overflow. The contribution in [22] considers the fact that the number of hidden terminal nodes changes when the distance between transmitter and receiver varies. Therefore, the PDR is expressed as a function of such distance. However, the study considers that there are no propagation errors up to a distance $R$ when there is no interference, and the interference has also a predefined range. The analytical model proposed in [23] is based on a two-dimensional Markov chain that models EDCA for the transmission of packets with different priorities. The proposal models the PDR and channel access delay for different traffic densities considering a fixed transmission range $R$. The work in [24] extends the model in [23] to model the mean, deviation and probability distribution of the channel access delay in IEEE 802.11p. The models in [21]-[24] consider the impact of hidden terminals but simplify the propagation and interference modeling, e.g., considering ideal propagation conditions up to a certain distance between transmitter and receiver.

The models presented in [25] and [26] consider the hidden terminal problem and propagation errors. However, they both consider a fixed communications range and propagation errors are modeled through a fixed error probability. This is an improvement compared to considering ideal propagation conditions but still cannot accurately model realistic V2X communications since in practice the range is not fixed and the error probability should depend on the distance between transmitter and receiver. In addition, these two models do not consider the received signal power of interfering packets. To the authors' knowledge, the analytical model that most realistically considers the propagation conditions was proposed in [27]. It models broadcast V2X communications under Rayleigh fading. The received signal power is modeled using an exponential random variable with certain mean that depends on the distance between transmitter and receiver. The receiver can decode the packet successfully if the received SINR (Signal to Interference and Noise Ratio) exceeds a threshold. However, the model from [27] does not consider the hidden terminal problem, which is an important limitation. Moreover, the derived metrics measure the average network performance and cannot quantify the V2X communications performance between two vehicles at a given distance.

The model proposed in [28] is one of the most complete models proposed to date to model the V2X communications performance using IEEE 802.11p. The model considers hidden terminals, includes propagation effects using a pathloss model and takes into account the received SINR to discard or not the packets that have collided due to a transmission by a hidden node. In addition, the model calculates the V2X communications performance as a function of the distance between transmitter and receiver. However, the model proposed in [28] still has relevant limitations. Considering a pathloss model to quantify the received signal strength is equivalent to considering ideal communications up to a certain distance $R$ in absence of packet collisions. At distances higher than $R$, the received signal is lower than a certain threshold and thus the packets cannot be received even without interferences. This results in a PDR equal to 1 for distances below $R$ and equal to 0 for distances higher than $R$ when there are no packet collisions. However, it is well-known that the received signal power is highly variable in vehicular networks. This variability results in that the PDR in absence of interference does not sharply decrease to zero from a certain distance, as demonstrated in multiple field tests such as the ones conducted by the CAMP in the US [29] or [30]. To demonstrate the importance of considering all these factors, the model proposed in this paper is compared with the model proposed in [28].

The conducted review has shown that despite the existence of previous IEEE 802.11p analytical models, there is yet the need for contributions that can more accurately model the performance of IEEE 802.11p-based V2X communications. This is the contribution of this paper that progresses the state-of-the-art by presenting a novel analytical model that accurately quantifies the PDR as a function of the distance between transmitter and receiver by jointly considering all relevant factors that influence the performance of V2X communications (including propagation, interference and hidden terminal). In addition, we propose new analytical models to quantify the probability of the four different types of packet errors in IEEE 802.11p, and the first analytical model that accurately estimates the CBR metric also under high channel load levels.

## III. OVERVIEW OF IEEE 802.11P

IEEE 802.11p adapts the IEEE 802.11a physical and MAC layers for vehicular communications. At the physical layer, IEEE 802.11p uses OFDM (Orthogonal Frequency Division Multiplexing) with a channel bandwidth of 10 MHz. IEEE 802.11p supports data rates ranging from 3 to 27 Mbps using convolutional coding rates 1/2, 2/3 or 3/4 and BPSK, QPSK, 16-QAM or 64-QAM modulations.

The basic MAC channel access method of IEEE 802.11p is CSMA/CA (Carrier Sense Multiple Access with Collision

Avoidance). Using CSMA/CA, a vehicle must sense the radio channel whenever it has a new packet ready to transmit. If it senses the channel as idle, it can start its transmission. If it senses the channel as busy because another vehicle is transmitting, it will defer its transmission until the channel becomes free again. In addition, when it senses the channel as busy, it sets randomly a backoff time that is a multiple of the parameter *aSlotTime*. The vehicle then decreases the backoff time every time it senses the channel as idle, and the vehicle starts its packet transmission when its backoff time reaches zero. This random backoff time helps reducing packet collisions when multiple vehicles are simultaneously deferring their transmissions. However, packet collisions cannot be completely avoided given the distributed nature of the IEEE 802.11p channel access mechanism for V2V communication. Two vehicles can still simultaneously transmit when their backoff time expires with a time difference below *aSlotTime*=13μs, i.e. nearly at the same time. When this happens, the two vehicles detect the channel as free and simultaneously start their transmission. These simultaneous transmissions are known as concurrent transmissions. Simultaneous transmissions can also occur due to the well-known hidden-terminal problem. The hidden-terminal problem occurs when two nodes cannot detect each other's transmissions (i.e. they are *hidden*), but their transmission ranges are not disjoint. Since they do not detect each other, these two nodes can transmit at the same time. When this happens, their transmissions are simultaneously received by nearby vehicles, experiencing a packet collision and a potential data loss due to the interferences.

## IV. Analytical models

IEEE 802.11p transmissions can encounter four different types of packet errors that are mutually exclusive. A packet error can only be classified within one of these four types:

- A packet is lost if it is received with a signal power below the sensing power threshold since the decoding process cannot be initiated. This error is referred to as SEN error.
- If the received signal power is above the sensing power threshold, a packet is lost if the receiver is busy decoding another packet. This error is referred to as RXB error.
- If the received signal power is higher than the sensing power threshold and the radio interface is free (i.e., it is not busy decoding another packet), a packet can be lost due to propagation effects. This error occurs if the SNR of a received packet is insufficient to successfully decode it. This type of error is referred to as PRO error.
- A packet can also be lost due to interference and packet collisions from other vehicles if its received signal power is higher than the sensing power threshold when the radio interface is free, but an interfering packet arrives while the packet is being decoded. These errors occur if the SINR is such that the packet cannot be successfully decoded and are referred to as COL errors. A packet with error is classified as COL if it is not classified as any of the previous errors.

A packet is correctly received if none of the identified types of error occur. The PDR can then be expressed as a function of the probability of each type of transmission error ($\delta_{SEN}$, $\delta_{RXB}$, $\delta_{PRO}$, and $\delta_{COL}$, respectively) as follows:

$$PDR(d_{t,r}) = (1-\delta_{SEN}(d_{t,r})) \cdot (1-\delta_{RXB}(d_{t,r})) \\ \cdot (1-\delta_{PRO}(d_{t,r})) \cdot (1-\delta_{COL}(d_{t,r})) \quad (1)$$

Alternatively, we can normalize the probability of each type of error and express the PDR as follows:

$$PDR(d_{t,r}) = 1 - \hat{\delta}_{SEN}(d_{t,r}) - \hat{\delta}_{RXB}(d_{t,r}) - \hat{\delta}_{PRO}(d_{t,r}) - \hat{\delta}_{COL}(d_{t,r}) \quad (2)$$

where

$$\hat{\delta}_{SEN}(d_{t,r}) = \delta_{SEN}(d_{t,r}) \quad (3)$$

$$\hat{\delta}_{RXB}(d_{t,r}) = (1-\delta_{SEN}(d_{t,r})) \cdot \delta_{RXB}(d_{t,r}) \quad (4)$$

$$\hat{\delta}_{PRO}(d_{t,r}) = (1-\delta_{SEN}(d_{t,r})) \cdot (1-\delta_{RXB}(d_{t,r})) \cdot \delta_{PRO}(d_{t,r}) \quad (5)$$

$$\hat{\delta}_{COL}(d_{t,r}) = (1-\delta_{SEN}(d_{t,r})) \cdot (1-\delta_{RXB}(d_{t,r})) \\ \cdot (1-\delta_{PRO}(d_{t,r})) \cdot \delta_{COL}(d_{t,r}) \quad (6)$$

$$0 \leq \delta_{SEN}, \delta_{RXB}, \delta_{PRO}, \delta_{COL} \leq 1 \quad (7)$$

$$0 \leq \hat{\delta}_{SEN} + \hat{\delta}_{RXB} + \hat{\delta}_{PRO} + \hat{\delta}_{COL} \leq 1 \quad (8)$$

Each of the error probabilities range between 0 and 1 (eq. (7)) and the sum of the normalized probabilities is lower or equal than one (eq. (8)). The PDR expression in (1) is obtained by substituting the normalized probabilities of eq. (3)-(6) in eq. (2).

To calculate the PDR, we first derive the probability of each type of error as a function of the distance $d_{t,r}$ between transmitter and receiver. Fig. 1 shows the flow chart of the main steps and equations used to calculate the different error probabilities. To calculate the different probabilities and the PDR, we consider a transmitting vehicle $v_t$ and a receiving vehicle $v_r$. Without loss of generality, we consider a highway scenario with a traffic density of $\beta$ vehicles per meter and a distance between vehicles of $1/\beta$ meters. All vehicles periodically transmit $\lambda$ packets per second with transmission power $P_t$. All packets have a payload size of $B$ bytes and are transmitted with data rate $DR$ on a 10 MHz channel at 5.9 GHz. Table II lists the main variables and parameters used to derive and describe the models.

### A. SEN errors

A SEN error occurs when the packet's received signal power is below the sensing power threshold $P_{SEN}$, since the decoding of the packet cannot be initiated. This type of error depends on the transmission power, the sensing power threshold, the propagation and the distance between transmitter and receiver, and it is not influenced by the channel access scheme at the MAC. We calculate the probability of SEN error following [31]. To this aim, we first compute the signal power $P_r$ at the receiver (in dB) as:

$$P_r(d_{t,r}) = P_t - PL(d_{t,r}) - SH \quad (9)$$

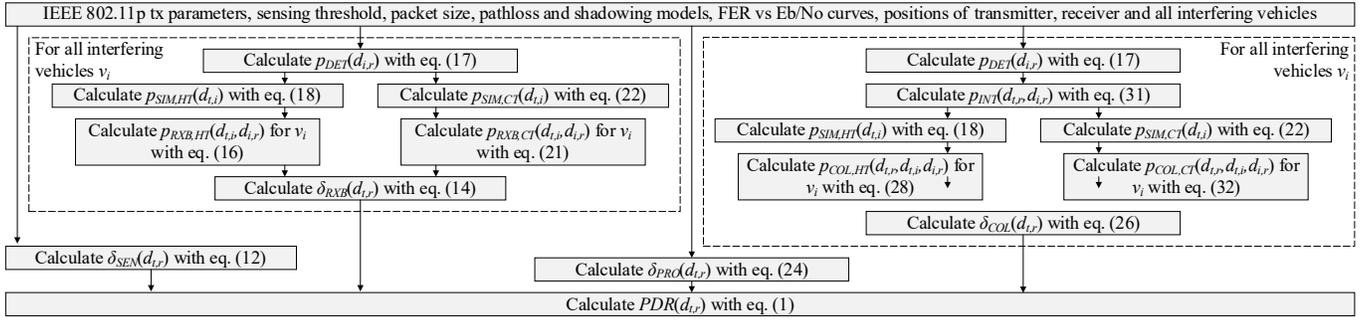

Fig. 1. Flow chart of the main steps and equations used to calculate the different packet error probabilities that are used to compute the PDR.

TABLE II. VARIABLES

| Variable | Description |
|---|---|
| $\beta$ | Traffic density (vehicles/meter) |
| $CBR$ | Channel Busy Ratio |
| $CBR_u$ | Upper bound of the Channel Busy Ratio |
| $\delta_{COL}$ | Probability of packet loss due to packet collision from any vehicle |
| $\delta^i_{COL}$ | Probability of packet loss due to packet collision from vehicle $v_i$ |
| $\delta^i_{COL,HT}$ | Probability of packet loss due to packet collision from vehicle $v_i$ due to the hidden terminal effect |
| $\delta^i_{COL,CT}$ | Probability of packet loss due to packet collision from vehicle $v_i$ due to concurrent transmission |
| $\delta_{PRO}$ | Probability of packet loss due to propagation effects |
| $\delta_{RXB}$ | Probability of packet loss because the receiver is busy |
| $\delta^i_{RXB}$ | Probability of packet loss because the receiver is busy due to a transmission from vehicle $v_i$ |
| $\delta_{SEN}$ | Probability of packet loss due to received signal below sensing threshold |
| $DR$ | Data rate |
| $p^i_{RXB,HT}$ | Probability of packet loss because the receiver is busy due to a transmission from vehicle $v_i$. Vehicles $v_i$ and $v_t$ simultaneously transmit because they are hidden. |
| $p^i_{RXB,CT}$ | Probability of packet loss because the receiver is busy due to a transmission from vehicle $v_i$. Vehicles $v_i$ and $v_t$ simultaneously transmit due to a concurrent transmission. |
| $\lambda$ | Packet transmission frequency (Hz) |
| $PDR$ | Packet Delivery Ratio |
| $PSR$ | Packet Sensing Ratio |
| $P_i$ | Received interference power from vehicle $v_i$ (dBm) |
| $P_r$ | Received power from vehicle $v_t$ (dBm) |
| $P_{SEN}$ | Sensing threshold (dBm) |
| $P_t$ | Transmission power (dBm) |
| $p_{DET}$ | Probability of detecting a packet |
| $p_{SINR}$ | Probability of packet loss due to low $SINR$ |
| $p_{SIM,HT}$ | Probability that $v_t$ and $v_i$ simultaneously transmit due to the hidden terminal problem |
| $p_{SIM,CT}$ | Probability that $v_t$ and $v_i$ simultaneously transmit due to concurrent transmission |
| $SINR$ | Signal-to-Interference-and-Noise Ratio (dB) |
| $SNR$ | Signal-to-Noise Ratio (dB) |

where $P_t$ is the transmission power, $PL(d_{t,r})$ is the pathloss at $d_{t,r}$, and $\sigma$ is the variance of the shadowing ($SH$) that is modeled as a log-normal random distribution with zero mean. The probability that $P_r$ is lower than the sensing threshold $P_{SEN}$ is:

$$\delta_{SEN}(d_{t,r}) = \int_{-\infty}^{P_{SEN}} f_{P_r,d_{t,r}}(p)dp \quad (10)$$

where $f_{P_r,d_{t,r}}(p)$ represents the PDF of $P_r$ at $d_{t,r}$. Since the shadowing follows a log-normal random distribution, we can express the PDF of $P_r$ as:

$$f_{P_r,d_{t,r}}(p) = \frac{1}{\sigma\sqrt{2\pi}} \exp\left(-\left(\frac{P_t - PL(d_{t,r}) - p}{\sigma\sqrt{2}}\right)^2\right) \quad (11)$$

and the probability that $P_r$ is lower than $P_{SEN}$ as:

$$\delta_{SEN}(d_{t,r}) = \frac{1}{2}\left(1 - erf\left(\frac{P_t - PL(d_{t,r}) - P_{SEN}}{\sigma\sqrt{2}}\right)\right) \quad (12)$$

where $erf$ is the well-known error function.

$1-\delta_{SEN}$ is the PSR (Packet Sensing Ratio). We can then derive an analytical expression of the PSR at any distance $d$ that we use to compute some of the other error probabilities:

$$PSR(d) = 1 - \delta_{SEN}(d) = \frac{1}{2}\left(1 + erf\left(\frac{P_t - PL(d) - P_{SEN}}{\sigma\sqrt{2}}\right)\right) \quad (13)$$

### B. RXB errors

A packet can be decoded if its received signal power is higher than the sensing threshold. However, the decoding process can only start if the radio interface is not busy receiving another packet. If it is busy, the packet cannot be received and we consider that the packet is lost. We refer to this type of error as an RXB error. RXB errors depend on the CSMA/CA channel access scheme but not on the received signal power of the discarded packet (and hence not on the transmission power, distance and propagation conditions). We should note that RXB errors exclude SEN errors.

An RXB error occurs when a packet transmitted by $v_t$ is received by $v_r$ with a received signal power higher than the sensing threshold, but the radio interface of $v_r$ is busy receiving another packet from any vehicle $v_i$. We denote $\delta^i_{RXB}(d_{t,r}, d_{t,i}, d_{i,r})$ as the probability of RXB error because $v_r$ is busy receiving a packet from vehicle $v_i$ when it receives a packet transmitted by $v_t$. This probability depends on the distance between $v_t$ and $v_r$, the distance between $v_t$ and $v_i$, and the distance between $v_t$ and $v_r$. We then calculate the probability that none of the interfering vehicles $v_i$ provokes an RXB error as $\prod_i(1 - \delta^i_{RXB}(d_{t,r}, d_{t,i}, d_{i,r}))$. Thus, the probability of RXB error is computed as the inverse of this probability:

$$\delta_{RXB}(d_{t,r}) = 1 - \prod_i\left(1 - \delta^i_{RXB}(d_{t,r}, d_{t,i}, d_{i,r})\right) \quad (14)$$

If vehicle $v_i$ is transmitting a packet, $v_t$ will start a transmission if it cannot detect the transmission of $v_i$ (hidden terminal problem), or if $v_t$ and $v_i$ finish their backoff time at same time (concurrent transmission). The probability of RXB error because the radio interface of $v_r$ is busy receiving another packet from a given vehicle $v_i$ while $v_t$ transmits a packet can then be expressed as:

$$\delta^i_{RXB}(d_{t,r}, d_{t,i}, d_{i,r}) = p^i_{RXB,HT}(d_{t,i}, d_{i,r}) + p^i_{RXB,CT}(d_{t,r}, d_{t,i}, d_{i,r}) \quad (15)$$

where $p^i_{RXB,HT}$ represents the probability of RXB error due to the hidden terminal problem and $p^i_{RXB,CT}$ represents the probability of RXB error due to a concurrent transmission. We calculate $p^i_{RXB,HT}$ as the multiplication of the probability $p_{SIM,HT}(d_{t,i})$ that vehicles $v_t$ and $v_i$ simultaneously transmit by the probability $p_{DET}(d_{i,r})$ that $v_r$ detects the packet transmitted by $v_i$ (i.e. the received signal power is higher than $P_{SEN}$):

$$p^i_{RXB,HT}(d_{t,i}, d_{i,r}) = p_{SIM,HT}(d_{t,i}) \cdot p_{DET}(d_{i,r}) \quad (16)$$

where $p_{DET}(d_{i,r})$ is the PSR at distance $d_{i,r}$ between $v_i$ and $v_r$:

$$p_{DET}(d_{i,r}) = PSR(d_{i,r}) \quad (17)$$

$p_{SIM,HT}(d_{t,i})$ is the probability that $v_t$ and $v_i$ simultaneously transmit considering that they are hidden and thus do not detect each other. It is calculated as:

$$p_{SIM,HT}(d_{t,i}) = \lambda \cdot T \cdot \frac{1 - PSR(d_{t,i})}{\Omega(d_{t,i})} \quad (18)$$

where $T$ is the packet duration, and $\lambda \cdot T$ represents the fraction of time consumed by the transmissions of vehicle $v_t$. The term $1 - PSR(d_{t,i})$ represents the probability that vehicle $v_t$ does not detect a packet transmitted by $v_i$, which is equal to the probability that vehicle $v_i$ does not detect a packet transmitted by $v_t$. $\Omega(d_{t,i})$ represents the fraction of time that both $v_t$ and $v_i$ detect the channel as free at the same time, and is calculated as:

$$\Omega(d_{t,i}) = 1 - CBR \cdot R_{PSR}(d_{t,i}) \quad (19)$$

where the *CBR* is the fraction of time that the channel is sensed as busy. We present in section IV.E a new analytical model of the CBR that is derived and validated in this study. In eq. (19), $CBR \cdot R_{PSR}(d_{t,i})$ is the fraction of time that both $v_t$ and $v_i$ detect the channel as busy at the same time, which depends on the distance $d_{t,i}$ between $v_t$ and $v_i$. $R_{PSR}(d_{t,i})$ is the autocorrelation of the PSR function at $d_{t,i}$. When $v_t$ and $v_i$ are close to each other, it is very likely that both detect the same transmissions from other vehicles. As a consequence, they detect the channel as busy at the same time and $R_{PSR}(d_{t,i}) = 1$. On the other hand, it is likely that $v_t$ and $v_i$ sense the channel as busy independently from each other when the distance between them is large. In this case, $R_{PSR}(d_{t,i}) = 0$. These properties are satisfied by the autocorrelation of the PSR function that is equal to:

$$R_{PSR}(d_{t,i}) = \sum_{j=-\infty}^{+\infty} PSR\left(\left|\frac{j}{\beta} + d_{t,i}\right|\right) \cdot PSR\left(\left|\frac{j}{\beta}\right|\right) \quad (20)$$

where the PSR is computed as a function of $j/\beta$ since we assume in this model that the distance between two consecutive vehicles is $1/\beta$ when the traffic density is $\beta$.

We use a similar method to calculate the probability $p^i_{RXB,CT}$ of RXB error due to a concurrent transmission. In this case, the packets transmitted by $v_t$ and $v_i$ arrive nearly at the same time to $v_r$. The proposed model assumes that the receiver starts decoding the packet received with higher power and discards the other one. Therefore, this type of error is only produced when $d_{i,r} < d_{t,r}$, and it can be expressed as:

$$p^i_{RXB,CT}(d_{t,r}, d_{t,i}, d_{i,r}) = \begin{cases} p_{SIM,CT}(d_{t,i}) \cdot p_{DET}(d_{i,r}) & \text{if } d_{i,t} < d_{t,r} \\ 0 & \text{if } d_{i,t} \geq d_{t,r} \end{cases} \quad (21)$$

As it can be observed, the probability $p^i_{RXB,CT}$ depends on the probability $p_{DET}(d_{i,r})$ that $v_r$ detects the transmission of $v_i$, which can be calculated with eq. (17). To calculate $p_{SIM,CT}(d_{t,i})$, we have to take into account that $v_t$ and $v_i$ can detect each other's transmissions, but they simultaneously transmit because they finish their respective backoff timer with a time difference below $\tau = aSlotTime$. $p_{SIM,CT}(d_{t,i})$ can then be calculated as:

$$p_{SIM,CT}(d_{t,i}) = \lambda \cdot \tau \cdot \frac{PSR(d_{t,i})}{\Omega(d_{t,i})} \quad (22)$$

*C. PRO errors*

A packet is decoded if its received signal power is higher than the sensing threshold and the radio interface is not decoding another packet. The packet can be lost due to a PRO error if its SNR is not sufficient for the receiver to successfully decode it. This type of error only quantifies errors due to propagation and does not consider interferences and packet collisions. As a result, PRO errors depend on the same factors as SEN errors (transmission power, sensing power threshold, propagation and distance between transmitter and receiver), and also on the data rate (i.e. on the modulation and coding scheme). The probability of not correctly receiving a packet due to propagation errors is denoted as $\delta_{PRO}$, and excludes the errors included in $\delta_{SEN}$ and $\delta_{RXB}$.

The probability of experiencing a PRO error depends on the PHY layer performance of the radio interface at the receiver. In this study, the PHY layer performance is modeled using Frame Error Rate (FER) curves as a function of the $E_b/N_0$ (or SNR per bit) from [32]. These curves were obtained considering a time-varying multipath channel. We model the *SNR* at a receiver as a random variable expressed in dB as:

$$SNR(d_{t,r}) = P_r(d_{t,r}) - N_0 = P_t - PL(d_{t,r}) - SH - N_0 \quad (23)$$

where $N_0$ is the noise power. The pathloss (*PL*) is constant for a given distance between transmitter and receiver. As a result, the SNR follows the same random distribution as the shadowing (*SH*) but with a mean value equal to $P_t - PL - N_0$. The probability $\delta_{PRO}$ that a packet is lost due to propagation effects is then:

$$\delta_{PRO}(d_{t,r}) = \sum_{s=-\infty}^{+\infty} FER(s) \cdot f_{E_b/N_0|P_r > P_{SEN}, d_{t,r}}(s) \quad (24)$$

where $FER(s)$ denotes the FER for $E_b/N_0 = s$ and

$$f_{E_b/N_0|P_r>P_{SEN},d_{t,r}}(s) = \begin{cases} \dfrac{f_{E_b/N_0,d_{t,r}}(s)}{1-\delta_{SEN}} & \text{if } P_r > P_{SEN} \\ 0 & \text{if } P_r \leq P_{SEN} \end{cases} \quad (25)$$

is the PDF of the $E_b/N_0$ experienced by vehicle $v_r$ at a distance $d_{t,r}$ for those $E_b/N_0$ values for which the $P_r$ is higher than the sensing threshold $P_{SEN}$. This PDF function therefore omits the packets that have been received with a signal power lower than the sensing power threshold $P_{SEN}$ since these packets have already been included in $\delta_{SEN}$ in eq. (12). $f_{E_b/N_0,d_{t,r}}(s)$ must be normalized by 1-$\delta_{SEN}$ in eq. (25) so that the probability $\delta_{PRO}$ of losing a packet due to propagation effects is between 0 and 1.

### D. COL errors

A packet that is being decoded (i.e., its received signal power is higher than the sensing threshold and is received when the radio interface is not decoding another packet) can also be lost due to the interference generated by other vehicles. This type of error is referred to as COL error, and occurs if the interference results in an SINR that prevents the successful decoding of the packet. Interference is generated by packet collisions with transmissions from other vehicles. The probability of packet collisions depends on the channel access scheme and the received signal power of the packet being decoded at the receiver and the interfering packets. The probability of losing a packet due to packet collisions is denoted as $\delta_{COL}$, and excludes the error types previously described.

To calculate $\delta_{COL}$, we consider that COL errors are produced when a packet transmitted by any interfering vehicle $v_i$ overlaps in time at a receiving vehicle $v_r$ with the packet transmitted by vehicle $v_t$, and the resulting interference prevents the correct reception of the packet at $v_r$. This probability depends on the distance between $v_t$ and $v_r$, the distance between $v_t$ and $v_i$, and the distance between $v_i$ and $v_r$. We compute this probability as the probability that any vehicle $v_i$ provokes a collision ($\delta_{COL}^i$) at $v_r$ with the following equation:

$$\delta_{COL}(d_{t,r}) = 1 - \prod_i \left(1 - \delta_{COL}^i(d_{t,r}, d_{t,i}, d_{i,r})\right) \quad (26)$$

Vehicle $v_i$ can provoke a packet collision due to the hidden terminal problem or due to a concurrent transmission. Therefore, the probability that a packet is lost due to a packet collision with vehicle $v_i$ can be calculated as:

$$\delta_{COL}^i(d_{t,r}, d_{t,i}, d_{i,r}) = p_{COL,HT}^i(d_{t,r}, d_{t,i}, d_{i,r}) + p_{COL,CT}^i(d_{t,r}, d_{t,i}, d_{i,r}) \quad (27)$$

$p_{COL,HT}^i$ is the probability that the packet is lost due to a collision resulting from the hidden terminal effect. This occurs when vehicles $v_t$ and $v_i$ do not detect each other and simultaneously transmit generating sufficient interference to provoke a packet loss at $v_r$. $p_{COL,HT}^i$ is calculated as:

$$p_{COL,HT}^i(d_{t,r}, d_{t,i}, d_{i,r}) = p_{SIM,HT}(d_{t,i}) \cdot p_{INT}(d_{t,r}, d_{i,r}) + p_{SIM,HT}(d_{t,i}) \cdot p_{INT}(d_{t,r}, d_{i,r}) \cdot (1 - p_{DET}(d_{i,r})) \quad (28)$$

Eq. (28) is the sum of two terms that differentiate when the packet transmitted by $v_t$ arrives to $v_r$ before or after the interfering packet transmitted by $v_i$, as illustrated in Fig. 2. The first term corresponds to the case when the packet transmitted by $v_t$ arrives first (Fig. 2a), and is equal to the multiplication of $p_{SIM,HT}(d_{t,i})$ and $p_{INT}(d_{t,r}, d_{i,r})$. $p_{SIM,HT}(d_{t,i})$ is the probability that $v_t$ and $v_i$ simultaneously transmit when they are hidden, and can be calculated with eq. (18). $p_{INT}(d_{t,r}, d_{i,r})$ represents the probability that the signal power of the packet transmitted by $v_i$ is sufficiently high to prevent the reception at $v_r$ of the packet transmitted by $v_t$. The second term in eq. (28) corresponds to the case when the packet transmitted by $v_i$ arrives first (Fig. 2b). It is analogous to the first term, but it is multiplied by the probability that $v_r$ does not detect the transmission of $v_i$. This term is added because if $v_r$ detects the packet transmitted by $v_i$, it would start decoding it. In this case, the radio interface of $v_r$ will be busy when the packet transmitted by $v_t$ arrives at $v_r$, and the packet from $v_t$ would be lost (RXB error).

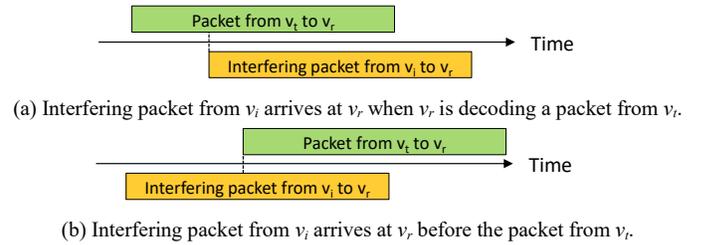

(a) Interfering packet from $v_i$ arrives at $v_r$ when $v_r$ is decoding a packet from $v_t$.

(b) Interfering packet from $v_i$ arrives at $v_r$ before the packet from $v_t$.

Fig. 2. Packet collision scenarios resulting from the hidden terminal effect.

To compute $p_{INT}(d_{t,r}, d_{i,r})$ in eq. (28), we consider that the interference received from vehicle $v_i$ over the received signal at $v_r$ is equivalent to additional noise. The *SINR* experienced by $v_r$ can then be computed in dB as:

$$SINR(d_{t,r}, d_{i,r}) = P_r(d_{t,r}) - P_i(d_{i,r}) \quad (29)$$

where $P_i$ is the signal power received at $v_r$ from $v_i$ plus the noise. In this context, *SINR* can be considered a random variable that is equal to the sum of two random variables ($P_r$ and $P_i$). The PDF of the *SINR* can thus be computed as the cross correlation of the PDF of $P_r$ and $P_i$ [33]. As a consequence, the probability that $v_r$ receives a packet with error due to low *SINR* is:

$$p_{SINR}(d_{t,r}, d_{i,r}) = \sum_{s=-\infty}^{+\infty} FER(s) \cdot f_{SINR|P_r>P_{SEN},d_{t,r},d_{i,r}}(s) \quad (30)$$

where the FER is again modeled using the FER curves as a function of the $E_b/N_0$ from [32]. Eq. (30) contains the packets that would have been lost even without the interference received from $v_i$. Since these packets were already considered when computing $\delta_{PRO}$ in eq. (24), we need to perform the following normalization:

$$p_{INT}(d_{t,r}, d_{i,r}) = \frac{p_{SINR}(d_{t,r}, d_{i,r}) - \delta_{PRO}(d_{t,r})}{1 - \delta_{PRO}(d_{t,r})} \quad (31)$$

In eq. (27), $p_{COL,CT}^i(d_{t,r}, d_{t,i}, d_{i,r})$ is the probability that the packet is lost at $v_r$ due to a collision because of a concurrent transmission. In this case, vehicles $v_t$ and $v_i$ could detect each other but they simultaneously transmit because their backoff timers finish nearly at the same time. As in eq. (21), we assume that the receiver starts decoding the packet received with higher power and discards the other one. Therefore, probability $p^i_{COL,CT}(d_{t,r}, d_{t,i}, d_{i,r})$ can be higher than zero only when $d_{i,r} \geq d_{t,r}$ so that the receiver starts decoding the packet from $v_t$ before the one from $v_i$ due to its higher received power level. Probability $p^i_{COL,CT}(d_{t,r}, d_{t,i}, d_{i,r})$ is then calculated for $d_{i,r} \geq d_{t,r}$ as the multiplication of the probability that $v_t$ and $v_i$ simultaneously transmit and the probability that the interference generated by $v_i$ is sufficient to provoke a packet loss:

$$p^i_{COL,CT}(d_{t,r}, d_{t,i}, d_{i,r}) = \begin{cases} p_{SIM,CT}(d_{t,i}) \cdot p_{INT}(d_{t,r}, d_{i,r}) & \text{if } d_{i,t} \geq d_{t,r} \\ 0 & \text{if } d_{i,t} < d_{t,r} \end{cases} \quad (32)$$

$p_{SIM,CT}(d_{t,i})$ is calculated with eq. (22) and $p_{INT}(d_{t,r}, d_{i,r})$ with eq. (31).

*E. CBR model*

This section presents a new analytical model to calculate the CBR. The CBR analytical model is used in this study to calculate the fraction of time that both $v_t$ and $v_i$ detect the channel as free at the same time with eq. (19). Therefore, it is used to calculate the probability of RXB and COL errors and the PDR. To calculate the CBR, we take into account the channel load generated by each vehicle. Each vehicle transmits $\lambda$ packets per second, and the duration of each packet is $T$. These packets only contribute to the CBR measured by a vehicle if the vehicle is able to detect them, i.e., if it receives them with a signal level above the sensing power threshold. Therefore, a vehicle located at a short distance from the transmitter would detect all these packets with probability $PSR(d)=1$. However, a vehicle at a large distance will most probably not detect any of them because $PSR(d)\approx 0$. Taking this effect into account, the channel load generated by a vehicle at a distance $d$ is:

$$load(d) = \lambda \cdot T \cdot PSR(d) \quad (33)$$

To calculate an upper bound of the CBR, we sum up the load generated by all vehicles $v_i$ using the following equation:

$$CBR_u = \sum_i load(d_i) = \sum_i load\left(\frac{i}{\beta}\right) = \beta \cdot \sum_i load(i) \quad (34)$$

where $d_i$ is the distance between vehicle $v_i$ and the ego vehicle that measures the CBR. Eq. (34) assumes that vehicle $v_1$ is at a distance $1/\beta$, vehicle $v_2$ is at $2/\beta$, etc. where $\beta$ is the traffic density in vehicles per meter. Eq. (34) considers the theory of the Riemann sum to take out the traffic density from the summation. $CBR_u$ is an upper bound of the actual CBR because it does not take into account packet collisions. When packets collide, they overlap in time and the channel is sensed as busy during a smaller amount of time. Such reduction is referred to as compression factor in [36], and significantly affects the CBR, especially for high channel load levels. To the authors' knowledge, there is no study that quantifies the compression factor to achieve an accurate estimation of the CBR for a wide range of input parameters. To solve this problem, we have conducted a regression analysis. In this analysis, we have used as input the CBR values obtained in all the simulations conducted in this study, and their corresponding analytical $CBR_u$ values computed with eq. (34). Simulations were needed to break the interrelation between CBR and packet collisions, since the probability of packet collision depends on the CBR and the CBR depends on the probability of packet collision. The regression analysis reveals that the best quadratic polynomial function that relates CBR and $CBR_u$ is:

$$CBR = p_1 \cdot CBR_u^2 + p_2 \cdot CBR_u + p_3 \quad (35)$$

with $p_1 = -0.2481$, $p_2 = 0.913$ and $p_3 = 0.003844$.

Fig. 3 plots the CBR as a function of $CBR_u$. In this figure, each dot represents the average CBR measured in one of the simulations conducted in this study. They include different traffic densities, transmission power levels, packet transmission frequencies, etc. The solid line represents the analytical model of the CBR proposed in eq. (35) and the dashed line represents the CBR model proposed in [36], which does not include the compression factor. As it can be observed, the proposed model matches with the simulations conducted. The highest CBR deviation observed between the proposed analytical model and the simulations is 0.02 and the mean absolute deviation is 0.006. Fig. 3 demonstrates the validity of the proposed CBR model under a wide range of parameters, and its higher accuracy compared to existing models, especially for high channel loads. The analytical model of the CBR proposed in eq. (35) has been used to obtain all the results presented in this paper. Similar results are obtained when $CBR_u$ is used as an approximation of the CBR for the scenarios considered. For high channel loads, the use of the analytical model of the CBR proposed in eq. (35) is recommended.

## V. VALIDATION

*A. Methodology*

The proposed analytical models have been implemented in Matlab and the source code is openly available at [5]. The models have been used to generate the analytical curves for the PDR and the probability of each transmission error as a function of the distance between transmitter and receiver for a wide set of input parameters. To validate the proposed models, we compare in section V.B the obtained analytical curves with the corresponding curves obtained by means of simulation. In addition, the proposed analytical model is compared in section

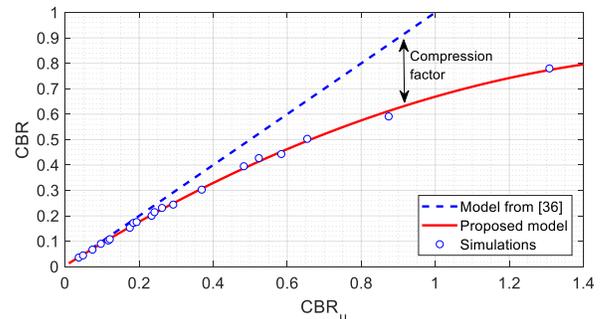

Fig. 3. CBR (Channel Busy Ratio) as a function of $CBR_u$.

V.C with the state-of-the-art model proposed in [28].

Simulations have been conducted using Veins to validate the models. Veins is an open-source framework for running vehicular network simulations that combines network and road traffic simulations using Omnet++ and SUMO (Simulation of Urban Mobility). SUMO is an open source, microscopic and multi-modal road traffic simulation tool widely used for research. The Winner+ B1 propagation model recommended in [34] for V2V communications has been implemented in Veins. The PHY layer performance of IEEE 802.11p is modelled using the link level LUTs (Look Up Tables) from [32] that express the FER vs. $E_b/N_0$.

By default, the comparison between the analytical models and the simulations is conducted considering that vehicles transmit packets with $B$=190 bytes at $\lambda$=10 Hz with a transmission power $P_t$=23 dBm and a data rate $DR$=6 Mbps (i.e using QPSK and a coding rate of ½). We have also validated the proposed model for different transmission power levels, packet transmission frequencies, data rates and packet sizes. The models have been validated for two different traffic densities (60 and 120 vehicles/km), which correspond to the Highway Fast and Highway Slow scenarios defined in [35]. Simulations are conducted in a highway scenario of 5 km with 4 lanes (2 lanes per driving direction) and a maximum speed of 70 km/h. The vehicles' mobility is controlled by SUMO using the Krauss car-following model. Vehicles are inserted in one edge of the highway and travel towards the other one. To avoid boundary effects, statistics are only taken from the vehicles located in the 2 km around the center of the simulation scenario. Table III summarizes the main parameters used in the validation.

To validate the proposed models, we have computed the PDR and the probabilities of the four identified errors in the simulator. To this aim, the simulator logs all the correctly received packets, and also logs and classifies all packets that are not correctly received into one of the four identified types of errors. For each packet, the logs also include the distance between transmitter and receiver.

To quantify the accuracy of the proposed analytical models, we compare the analytical curves and the curves obtained by simulation using the Mean Absolute Deviation (MAD) metric. This metric quantifies the absolute difference between two vectors of $M$ elements ($m_s$ and $m_a$), and is calculated as:

$$MAD[\%] = \frac{100}{M} \sum_{i=1}^{M} |m_s(i) - m_a(i)| \quad (36)$$

TABLE III. PARAMETERS

| Parameter | Values analyzed |
|---|---|
| Traffic density ($\beta$) | 60, 120 veh/km |
| Max. speed of vehicles | 70 km/h |
| Highway length | 5 km |
| Number of lanes | 4 (2 per direction) |
| Channel bandwidth | 10 MHz |
| Transmission power ($P_t$) | 15, 23, 30 dBm |
| Packet tx frequency ($\lambda$) | 10, 25 Hz |
| Packet size ($B$) | 190, 500 bytes |
| Data rate ($DR$) | 6, 18, 27 Mbps |

*B. Evaluation*

Fig. 4 compares the PDR obtained with the proposed analytical model (dashed lines) and with simulations (solid lines). The figure analyzes the impact of the data rate considering $P_t$=23 dBm and $B$=190 Bytes. The data rate affects the packet duration, and thus the channel load and the interference generated by each vehicle. Fig. 4a considers a traffic density of $\beta$=60 veh/km and a packet transmission frequency of $\lambda$=10 Hz, which results in low channel load levels (the CBR is below 10%). Fig. 4a shows that the PDR obtained with our analytical model closely matches the PDR obtained by simulation for all data rates evaluated. We should highlight the accuracy of our analytical model since simulations implement in detail IEEE 802.11p and realistically model the traffic mobility. Fig. 4b compares the analytical and simulation PDRs for a scenario with $\beta$=120 veh/km and $\lambda$=25 Hz that result in CBR levels between 16% and 44%. Fig. 4b shows that our model is also very accurate when the channel load increases. Fig. 4b shows only minor deviations for $DR$=6 Mbps, which corresponds to the highest load in this figure (CBR of 44%). This deviation is due to a small deviation in the modeling of the probability of packet loss due to collisions ($\delta_{COL}$) that is visible in Fig. 5. Fig. 5 compares the probability of each type of transmission error obtained with our analytical models and with simulations for $\beta$=120 veh/km and $\lambda$=25 Hz. The figure shows almost a perfect match for the probabilities of SEN, PRO and RXB errors, and only a small deviation for the COL error. However, we should note that such deviation is only present under the highest CBR levels since the analytical and simulation probabilities of COL error perfectly match when $\beta$=60 veh/km and $\lambda$=10 Hz.

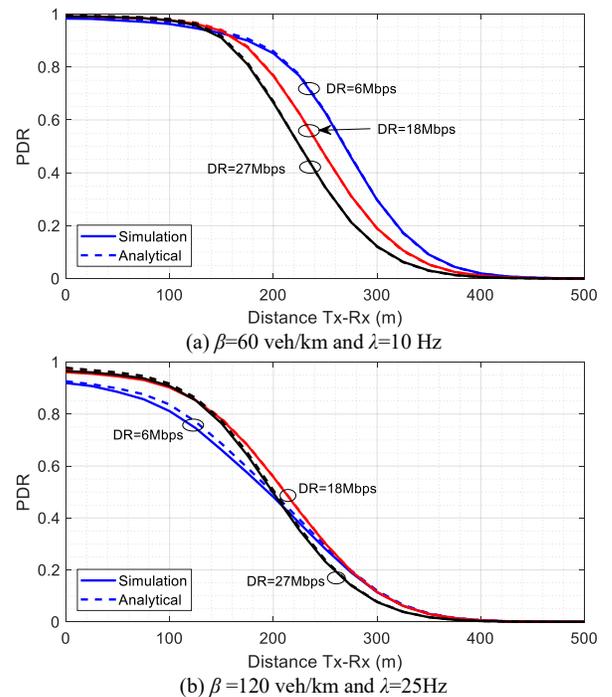

(a) $\beta$=60 veh/km and $\lambda$=10 Hz

(b) $\beta$=120 veh/km and $\lambda$=25Hz

Fig. 4. PDR as a function of the distance between transmitter and receiver. Pt=23dBm and B=190 bytes.

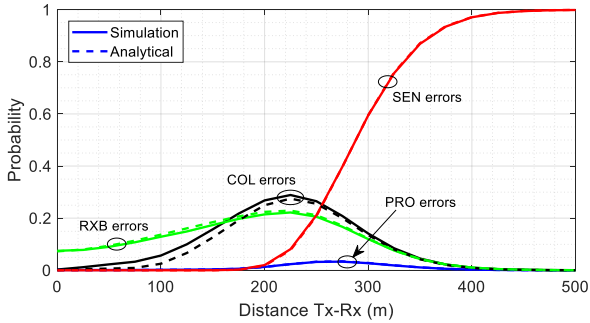

Fig. 5. Error probabilities as a function of the distance between transmitter and receiver. DR=6Mbps, Pt=23dBm, B=190 bytes, $\beta$=120 veh/km, $\lambda$=25Hz.

The effect of the transmission power on the accuracy of the analytical PDR model is analyzed in Fig. 6 for *DR*=6 Mbps and *B*=190 bytes. The transmission power affects the sensing and communications ranges. As a result, it influences the channel load and the number of vehicles that can generate a packet collision due to the hidden terminal problem or concurrent transmissions. In this case, Fig. 6a plots the results for a low channel load scenario ($\beta$=60 veh/km and $\lambda$=10 Hz) and Fig. 6b for a high channel load scenario ($\beta$=120 veh/km and $\lambda$=25 Hz). Fig. 6a shows again how the analytical PDR curves closely match the PDR curves obtained by simulation for low channel load levels (CBR below 15% in Fig. 6a). When the channel load increases, the analytical PDR curves closely follow the curves obtained by simulation. Fig. 6 shows that the analytical model is able to adequately capture the variation of the shape of the PDR as a function of the distance independently of the transmission power. Fig. 7 shows again that the analytical models for the probability of each type of transmission error closely match the probabilities obtained through simulations for different transmission power levels. Fig. 7 corresponds to the highest traffic density scenario since Fig. 6a shows that there is an almost perfect PDR match for the lowest density, and hence for the probabilities for all types of transmission errors.

Fig. 8 analyzes the accuracy of the proposed PDR analytical model for different packet sizes considering *DR*=6 Mbps and *Pt*=23 dBm. Increasing the packet size augments the channel load and the probability of packet collisions. Fig. 8 shows that the proposed analytical PDR models accurately matches the simulation PDR even if we increase the packet size.

Tables IV and V quantify the accuracy of the proposed analytical models. The tables report the MAD metric for the PDR and the four possible transmission errors in IEEE 802.11p under different conditions. The MAD metric measures the average deviation between the results obtained analytically and through simulations. The tables report in the last column the CBR level (analytically estimated with eq. (35)) for each combination of parameters in the tables. Table IV analyzes the impact of the DR on the accuracy of the proposed models. The table shows that the MAD metric for the PDR is below 1% for all cases. The transmission errors are also accurately modelled with their MAD metric below 1% for almost all errors and scenarios. Table V analyzes the impact of the transmission power on the MAD metric. This table shows again that the MAD for the PDR is below 1% in nearly all the configurations. It is only higher than 1% for the scenario with the highest channel load. In this scenario, the MAD for the PDR was below 3% even if the CBR was 59%. Results in Tables IV and V demonstrate again the high accuracy of the proposed analytical models. Only relatively small differences are observed in configurations that generate high CBR levels that will not likely be experienced in practical deployments thanks to congestion control protocols.

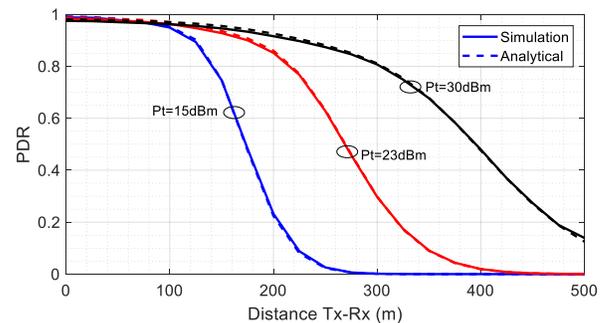

(a) $\beta$=60 veh/km and $\lambda$=10 Hz

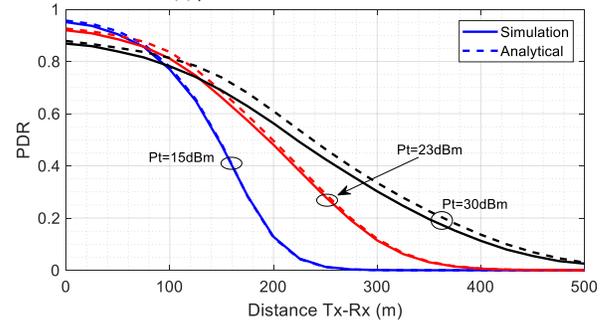

(b) $\beta$=120 veh/km and $\lambda$=25Hz

Fig. 6. PDR as a function of the distance between transmitter and receiver. DR=6Mbps and B=190 bytes.

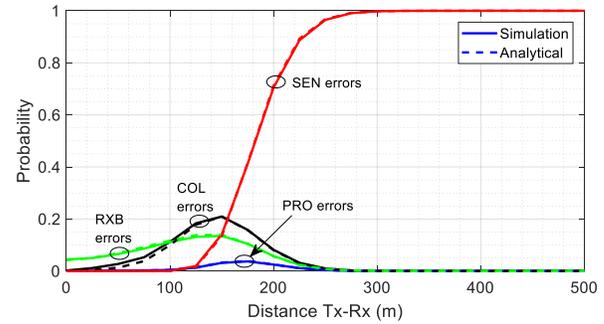

(a) Pt = 15dBm

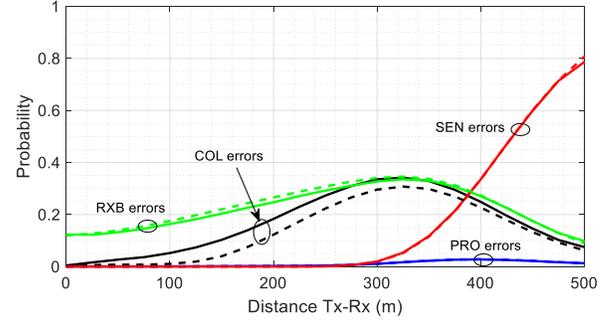

(b) Pt = 30dBm

Fig. 7. Error probability as a function of the distance between transmitter and receiver for $\beta$=120 veh/km, $\lambda$=25Hz, DR=6Mbps, B=190 bytes.

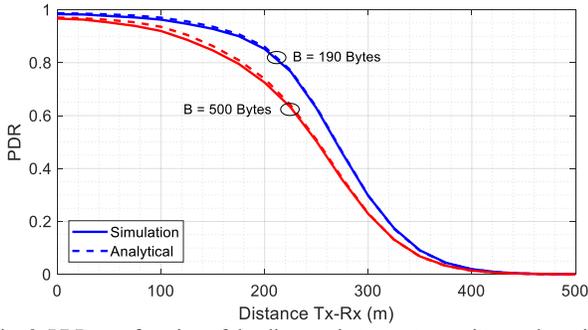

Fig. 8. PDR as a function of the distance between transmitter and receiver. $\beta$=60 veh/km $\lambda$=10 Hz, DR=6Mbps and Pt=23dBm.

TABLE IV. MAD FOR THE PDR AND THE DIFFERENT TYPES OF ERRORS. $Pt$=23DBM AND $B$=190 BYTES

| $\beta$ | $\lambda$ | DR | PDR | $\hat{\delta}_{SEN}$ | $\hat{\delta}_{RXB}$ | $\hat{\delta}_{PRO}$ | $\hat{\delta}_{COL}$ | CBR |
|---|---|---|---|---|---|---|---|---|
| 0.06 | 10 | 6 | 0.37 | 0.09 | 0.20 | 0.03 | 0.14 | 0.10 |
| | | 18 | 0.18 | 0.11 | 0.11 | 0.12 | 0.05 | 0.04 |
| | | 27 | 0.24 | 0.09 | 0.15 | 0.16 | 0.12 | 0.03 |
| 0.12 | 25 | 6 | 0.94 | 0.10 | 0.42 | 0.02 | 1.13 | 0.44 |
| | | 18 | 0.27 | 0.10 | 0.14 | 0.09 | 0.18 | 0.21 |
| | | 27 | 0.52 | 0.10 | 0.28 | 0.16 | 0.27 | 0.16 |

TABLE V. MAD FOR THE PDR AND THE DIFFERENT TYPES OF ERRORS. $DR$=6MBPS AND $B$=190 BYTES

| $\beta$ | $\lambda$ | $P_t$ | PDR | $\hat{\delta}_{SEN}$ | $\hat{\delta}_{RXB}$ | $\hat{\delta}_{PRO}$ | $\hat{\delta}_{COL}$ | CBR |
|---|---|---|---|---|---|---|---|---|
| 0.06 | 10 | 15 | 0.26 | 0.20 | 0.06 | 0.03 | 0.06 | 0.06 |
| | | 23 | 0.37 | 0.09 | 0.20 | 0.03 | 0.14 | 0.10 |
| | | 30 | 0.60 | 0.12 | 0.23 | 0.04 | 0.35 | 0.15 |
| 0.12 | 25 | 15 | 0.31 | 0.15 | 0.16 | 0.02 | 0.29 | 0.29 |
| | | 23 | 0.94 | 0.10 | 0.42 | 0.02 | 1.13 | 0.44 |
| | | 30 | 2.74 | 0.15 | 1.01 | 0.03 | 3.33 | 0.59 |

## C. Comparison

Fig. 9 compares the PDR obtained with the proposed model and the PDR obtained with the model proposed in [28]. For a fair comparison, the pathloss model, the sensitivity and receiving thresholds, the transmission parameters and the packet size in the model proposed in [28] have been adapted to match the ones used in this paper. Fig. 9a considers a low channel load scenario and Fig. 9b a high channel load scenario.

The figure shows that there are significant differences between the PDR curves obtained with the two models. The PDR of the model proposed in [28] sharply goes down to zero at the distance at which the received signal power is below the power needed to successfully decode the packet. This sharp decrease of the PDR from a certain distance does not match the simulation results and does not realistically represent the variation of the performance of V2X communications with the distance between transmitter and receiver [29][30]. This variation is better captured with our proposed model that models the variability of the received signal power with the distance thanks to the inclusion of the pathloss, shadowing and multipath fading effects. To highlight the impact of these effects, Fig. 9 also represents the PDR obtained with our model removing the multipath effect (deactivating in our model the LUTs of the FER curves) and the shadowing effect. Fig. 9a shows that similar PDR curves can be achieved with our model and the model from [29] under low channel load levels when the multipath and shadowing effects are omitted. The differences between the two PDR curves are due to the fact that the model in [28] does not take into account the received signal power of interfering packets for concurrent transmissions (i.e., it only considers it for hidden nodes) while our model does take it into account. The impact of these interfering packets might not be significant under low channel loads, but increases with the channel load and results in additional packet collisions that need to be captured by the models. This explains why the differences between the PDR obtained with the model in [28] and our proposed model (without shadowing and multipath fading) increases with the channel load (Fig. 9b). Fig. 9 also shows that intermediate results are obtained when only the multipath fading effect is removed from our model.

The results presented in this section clearly show the importance of adequately modeling the propagation and interference conditions to quantify the V2X communications performance. They also demonstrate the impact of some of the main features of the proposed model that are not available in most of the existing analytical models.

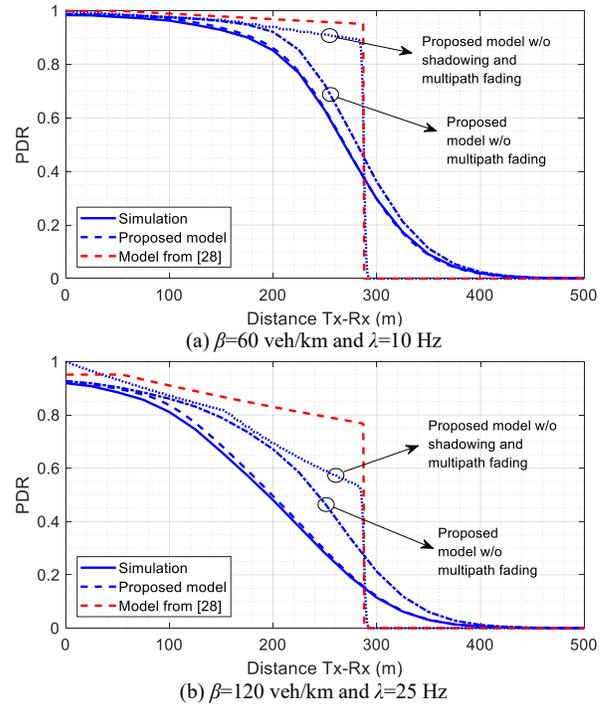

Fig. 9. PDR as a function of the distance between transmitter and receiver for our model and the model in [28]. DR=6Mbps, Pt=23dBm, B=190 bytes.

## VI. CONCLUSIONS

This paper proposes and validates an analytical model of the PDR that two vehicles using IEEE 802.11p experience as a function of their distance. The proposed model takes into account in detail the propagation and interference effects as well as the hidden terminal problem. We derive the PDR by analytically quantifying the probability of the four different types of possible transmission errors. The proposed models have been validated by means of simulation for a wide range of communication and traffic parameters. The obtained results demonstrate the high accuracy of the proposed models with a mean absolute deviation below 1% for most of the configurations analyzed. The validation demonstrates that the proposed models can be used under a wide range of conditions to accurately estimate the performance of IEEE802.11p-based V2X communications. The paper also presents and validates an analytical model of the CBR. The novelty of the proposed CBR model is that it is the first, to the authors' knowledge, that quantifies the compression factor to achieve an accurate estimation of the CBR for a wide range of input parameters. To facilitate their use by the community, we provide with this paper an open-source implementation of the proposed models.